\documentclass[aps,prl,twocolumn,showpacs,superscriptaddress,groupedaddress,superscriptaddress]{revtex4-1}  
\usepackage{graphicx}  
\usepackage{dcolumn}   
\usepackage{bm}        
\usepackage{amssymb}   
\usepackage{multirow}
\usepackage[usenames]{color}

\begin{document}

\title{
Comprehensive study on band-gap variations in $sp^3$-bonded semiconductors: roles of electronic states floating in internal space
}

\author{Yu-ichiro Matsushita}
\affiliation{Department of Applied Physics, The University of Tokyo, 
Tokyo 113-8656, Japan}
\author{Atsushi Oshiyama}
\affiliation{Department of Applied Physics, The University of Tokyo, 
Tokyo 113-8656, Japan}

\date{\today}

\begin{abstract}
We have performed electronic structure calculations to explore the band-gap dependence on polytypes for $sp^3$-bonded semiconducting materials, i.e., SiC, AlN, BN, GaN, Si, and diamond. In this comprehensive study, we have found that band-gap variation depending on polytypes is common in $sp^3$-bonded semiconductors; SiC, AlN, and BN exhibit smallest band gaps in $3C$ structure, whereas diamond does in $2H$ structure. We have also clarified that the microscopic mechanism of the band-gap variations is attributed to peculiar electron states $floating$ in internal channel space at the conduction-band minimum (CBM), and that internal channel length and the electro-static potential in channel affect the energy level of CBM.
\end{abstract}

\maketitle

\section{Introduction}

The development of the modern society has been mostly attributed to semiconductor technology. In most semiconductors from elemental to compound, each atom forms $sp^3$ bonds and take four-fold coordinated tetrahedral structure. It is well known that the crystal structure consisting of $sp^3$ bonds exhibit hundreds of polytypes \cite{Dissertation10}. Their structural difference is in the stacking of the tetrahedral units along the $\left<111\right>$ direction in cubic structure, and $\left<0001\right>$ direction in hexagonal structure. Zincblende and wurtzite structures are the most famous examples of them. The Zincblende structure is represented by the stacking sequence of ABC and the wurtzite by AB. Nomenclature adopted vastly is introduced here: each polytype is labeled by the periodicity of the stacking sequence $n$ and the symmetry (cubic or hexagonal) such as $2H$ (wurtzite), $3C$ (zinblende), $4H$, $6H$, etc.

These structural differences have been assumed to be minor in the electronic properties. It is because there are no differences in local atomic structure up to the 2nd nearest neighbor. Valence bands consist of $sp^3$-bonding orbitals and conduction bands $sp^3$-antibonding. However, it is reported that the stacking sequence affects electronic properties considerably in silicon carbide (SiC) \cite{Harris,kimoto}. SiC is indeed a manifestation of the polytypes: Dozens of polytypes of SiC are observed.  Yet, surprisingly, the band gaps vary by 40 \%, from 2.3 eV in 3C to 3.3 eV in 2H despite that the structures are locally identical to each other in all the polytypes \cite{Harris}. This phenomenon was difficult to be understood in conventional chemical pictures.

\begin{figure}
\begin{center}
\includegraphics[width=0.45\textwidth]{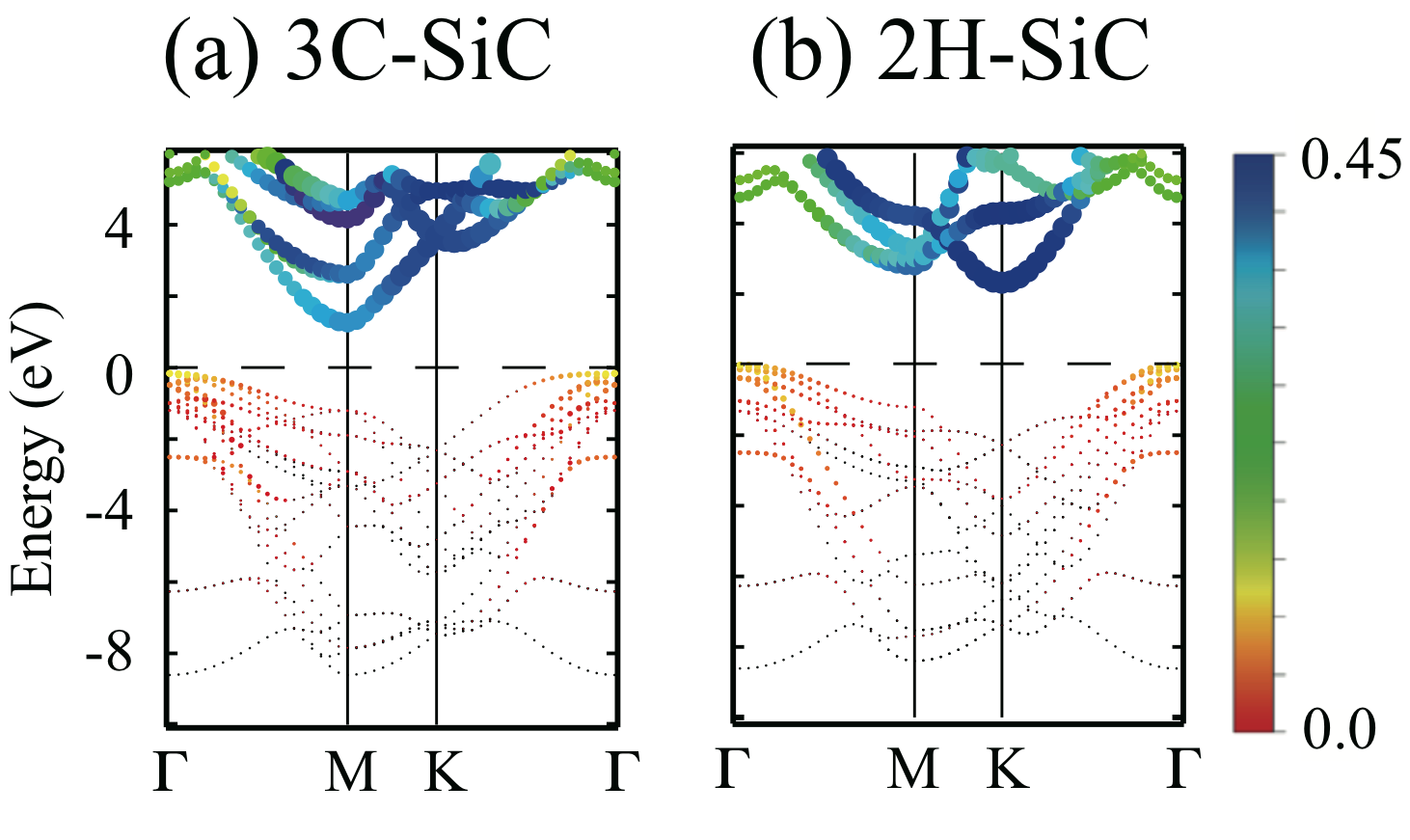}  
\end{center}
\caption{(Color online) Residual norms of the wavefunctions of the energy bands of 3$C$-SiC (a) and 2$H$-SiC (b). The residual norms are represented by the color and the size of the dots. The energy of the valence-band top is set to be 0. The residual norm which is a measure of the floating nature is calculated in the following procedure: From the pseudo-atomic orbitals $\{\phi_i^{\rm isolated}\}$ of isolated silicon and carbon atoms, we have composed orthonormal basis set $\{\phi_i^{\rm atom}\}$ with the Gram-Schmidt orthonormalization. Then we have calculated the squared residual norm, which is defines as $\left| \left|\phi_{n{\bf k}}\right> -\sum_i \left|\phi_i^{\rm atom}\right>\left<\phi_i^{\rm atom}|\phi_{n{\bf k}}\right> \right|^2$ for each band n at ${\bf k}$ point.}
\label{seibun}
\end{figure}

\begin{figure}
\begin{center}
\includegraphics[width=0.3\textwidth]{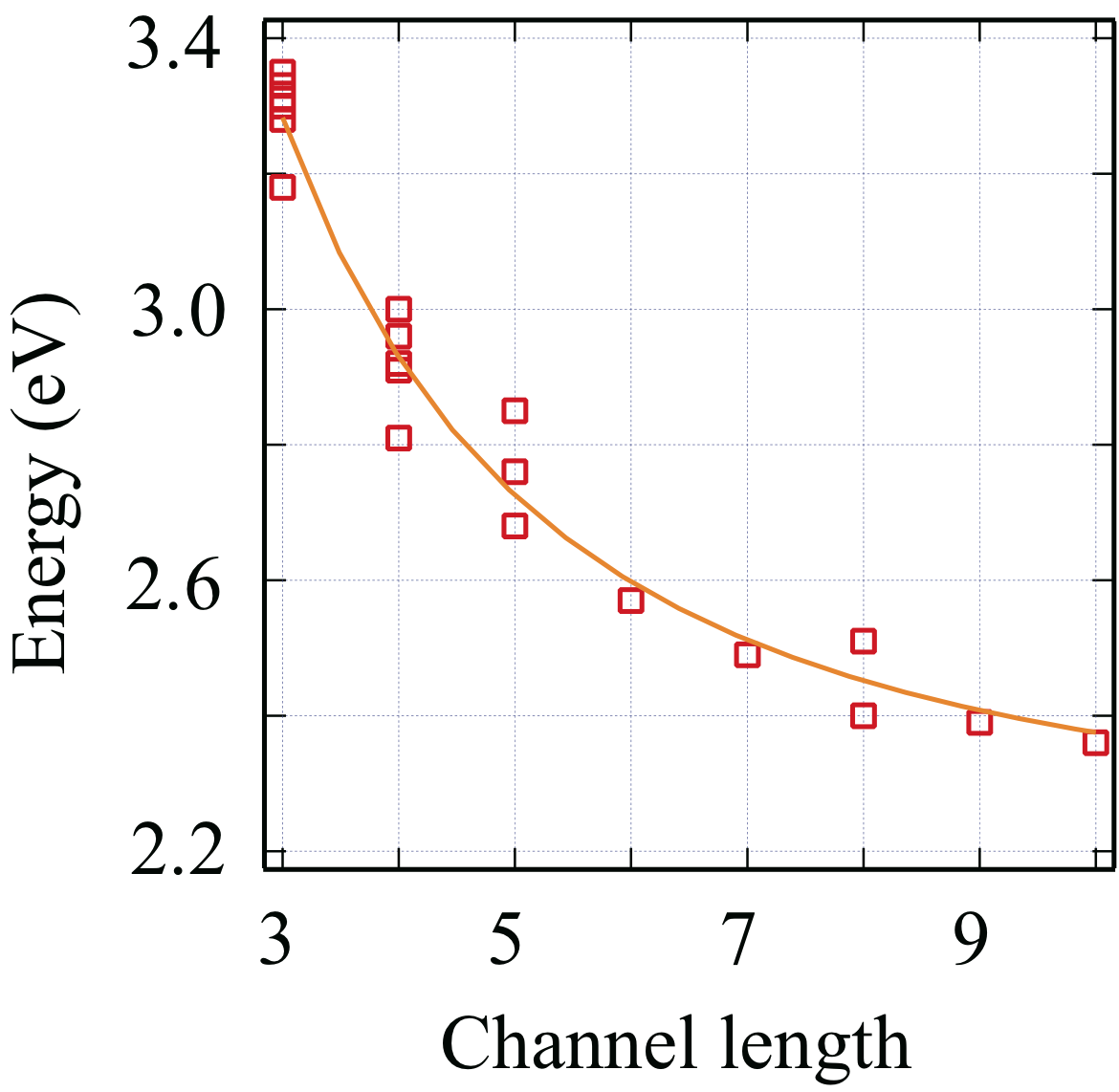}  
\end{center}
\caption{(Color online) Calculated band gaps as a function of channel length for 24 SiC polytypes with Heyd-Scuseria-Ernzerhof (HSE) functional \cite{HSE, HSE04, HSE05, HSE06, matsushita4}. The channel length is defined as the number of bilayers along the longest interstitial channels. The orange curve represents a fitting function of $y=2.21+21.85/(x+1.51)^2$. Specific values of each plot are shown in Appendix.}
\label{bandgap_HSE}
\end{figure}

We have recently reported \cite{matsushita1, matsushita2, matsushita3} the microscopic mechanism of band-gap variations in SiC polytypes based on the density functional theory (DFT) \cite{K-S, H-K}. It is found that continuum states exist in conduction bands of $sp^3$-bonded materials, and furthermore, such peculiar electron state appears at the conduction-band minima (CBM) in SiC polytypes. The wavefunction at the CBM is not distributed near atomic sites, but extends (or $floats$) in interstitial channels without atomic orbital character. This fact is clearly shown in Fig.~\ref{seibun}. The figure shows the calculated residual norms of the wavefunction of 3$C$-SiC after projecting it to the $s$- and $p$-atomic orbitals. While the wavefunction in valence bands is well described by $s-$ and $p-$atomic orbitals, conduction electrons cannot. This $floating$ character at the CBM is an important key to reveal the microscopic mechanism of the band-gap variations. 
Channel length changes depending on polytypes: $3C$ structure has infinite channel length along the $\left<110\right>$ direction, while similar channel structure is seen also in $6H$ structure with the length of about $7 a_0 / 2 \sqrt{2}$ along $\langle \bar{2}201 \rangle$ which is slanted relative to $\langle 0001\rangle$ direction with $a_0$ a lattice constant. $Floating$ states at the CBM extends in the internal space. Therefore, channel length and channel shapes are decisive in the positions of the CBM in energy space. In fact, we have found that the energy level of the floating state strongly depends on channel length via quantum confinement effect (see Fig.~\ref{bandgap_HSE}) \cite{matsushita2}. 



We have already clarified the microscopic mechanism of band-gap variations in SiC cases as shown in Fig.~\ref{seibun} and Fig.~\ref{bandgap_HSE}. From the similarity of the crystal structures, however, it is expected that the similar phenomena can be seen also in other $sp^3$-bonded semiconductors. In this study, we have investigated the possibility of band-gap variations in wide range of $sp^3$-bonded semiconductors, the effects of each atomic character on the electronic properties, and the differences between elemental and compounds semiconductors for Si, diamond, AlN, BN, and GaN. 

\section{ Calculation conditions}
Total-energy band-structure calculations were performed based on the DFT \cite{K-S, H-K} in this study using the plane-wave-basis-set $ab$ $initio$ program package, TAPP \cite{sugino,yamauchi,kageshima}. Our calculations have been performed in the generalized gradient approximations (GGA) \cite{PBE, perdew96}.  Nuclei and core electrons are simulated by either norm-conserving \cite{troullier} pseudo-potentials in the TAPP code. We generate norm-conserving pseudo-potential to simulate nuclei and core electrons, following a recipe by Troullier and Matins \cite{troullier}. The core radius $r_c$ is an essential parameter to determine transferability of the generated pseudo-potential. We have examined $r_c$ dependence of the calculated structural properties of benchmark materials and adopted the pseudo-potentials generated with the following core radii in this paper: 0.85 {\AA} for Si 3$s$, and 1.16 {\AA} for Si 3$p$, 1.06 {\AA} for Ga 4$s$ and 4$p$, and  1.48 {\AA} for Ga 4$d$,  0.64 {\AA} for N 2$s$ and 2$p$, 0.85 {\AA} for C 2$s$ and 2$p$, 1.06 {\AA} for Al 3$s$, 3$p$, and 3$d$,  0.847 {\AA} for B 2$s$ and 2$p$. 

\section{Results and discussion}
We first present our calculated band gaps for structurally optimized polytypes in the next subsection. In the following subsections, we describe the floating states in each polytype and the roles of floating states on band gap variations. 

\subsection{Optimized structures and their band gaps}
\begin{table*} 
\begin{center}
\caption{
Calculated hexagonal lattice constant $a$ and the ratio $c/na$ for different polytypes labeled as either $nH$ or $nC$ ($n=3$) of the various $sp^3$-bonded semiconductors. Calculated total energies per formula unit are also shown. The values are relative to the energy of the corresponding the most stable structure.}
\label{structural parameter}
\scalebox{1.0}[0.9]{
\begin{tabular}{cccccccc}
\hline
\multirow{2}{*}{Materials} &  \multicolumn{2}{c}{$a$ [\AA]} & \multicolumn{2}{c}{$c/na$} &\multirow{2}{*}{ $\Delta$E [meV]}
\\
                  &this work& Expt.                  &this work & Expt.                &         
\\ [2pt]  \hline  \\ [-4pt]
$2H$-SiC & 3.085 & 3.076 (Ref. \cite{2H-SiC})& 0.8217 & 0.8205 (Ref. \cite{2H-SiC}) & 7.1
\\
$3C$-SiC & 3.091 & 3.083 (Ref. \cite{3C-SiC}) & 0.8165 & 0.8165 (Ref. \cite{3C-SiC})& 1.2
\\
$6H$-SiC & 3.091 & 3.081 (Ref. \cite{6H-SiC})& 0.8180 & 0.8179 (Ref. \cite{6H-SiC})& 0 
\\ [2pt]  \hline  \\ [-4pt]
$2H$-AlN & 3.117 & 3.110 (Ref. \cite{2H-AlN}) & 0.8103 & 0.8005 (Ref. \cite{2H-AlN})& 0 
\\
$3C$-AlN & 3.112 & 3.090 (Ref. \cite{3C-AlN})& 0.8165 & 0.8165 (Ref. \cite{3C-AlN})&41.9
\\
$6H$-AlN & 3.112 & $-$ & 0.8148 & $-$ &28.4
\\ [2pt]  \hline  \\ [-4pt]
$2H$-BN & 2.556 & 2.553 (Ref. \cite{2H-BN})& 0.8252 & 0.8265 (Ref. \cite{2H-BN})& 35.5 
\\
$3C$-BN & 2.561 & 2.557 (Ref. \cite{2H-BN})& 0.8165 & 0.8165 (Ref. \cite{2H-BN})& 0 
\\
$6H$-BN & 2.556 & 2.500 (Ref. \cite{6H-BN})& 0.8203 & 0.8293 (Ref. \cite{6H-BN})& 9.6 
\\ [2pt]  \hline  \\ [-4pt]
$2H$-GaN & 3.255 & 3.189 (Ref. \cite{2H-GaN})& 0.8156 & 0.8130 (Ref. \cite{2H-GaN})& 0
\\
$3C$-GaN & 3.255 & 3.175 (Ref. \cite{3C-GaN})& 0.8165 & 0.8165 (Ref. \cite{3C-GaN})&15.3
\\
$6H$-GaN & 3.255 & $-$ & 0.8159 & $-$ & 9.5
\\ [2pt]  \hline  \\ [-4pt]
$2H$-Si & 3.853 & $-$ & 0.8238 & $-$ & 22.3
\\
$3C$-Si & 3.863 & 3.863 (Ref. \cite{3C-GaN})& 0.8165 & 0.8165 (Ref. \cite{3C-GaN})& 0
\\
$6H$-Si & 3.858 & $-$ & 0.8193 & $-$ & 4.0 
\\ [2pt]  \hline  \\ [-4pt]
$2H$-C & 2.503 & $-$ & 0.832 & $-$ & 50.9 
\\
$3C$-C & 2.514 & 2.519 (Ref. \cite{3C-GaN})& 0.8165 & 0.8165 (Ref. \cite{3C-GaN})& 0 
\\
$6H$-C & 2.508 & $-$ & 0.8230 & $-$ & 12.8 \\
\hline
\end{tabular}
}
\end{center}
\end{table*}

First, we have theoretically determined lattice constants in the hexagonal plane and along the stacking direction, $a$ and $c$ in the GGA. The obtained $a$ and the ratio $c/na$ of each polytype are listed in Table~\ref{structural parameter}, where $n$ represents the periodicity of stacking bilayers. The differences of $c/na$ among the polytypes are found to be extremely small. This fact means that the distortion along the c-axis is quite small. Our calculated lattice constants agree with available experimental data with an error of at most 2 \%. Table~\ref{structural parameter} also shows the calculated total energy differences ($\Delta E$) among the geometry-optimized polytypes. The table includes some polytypes not observed yet, .e.g, 6$H$-AlN. 
Yet, it is likely that these polytypes are synthesizable since the total energy difference is small, being in the range of 50 meV or less per molecular unit. 

The most energetically favorable polytype in SiC is the $6H$ followed by the $3C$ with the energy increase of 1.2~meV per SiC molecular unit. 
It is said that $4H$ is also one of the most energetically favorable polytypes \cite{Kobayashi}. Yet, the $6H$ structure is an often observed polytype in experiments, and our calculations show quite small difference in total energy than that of $4H$ polytype by 0.1 meV. Therefore, we discuss the $6H$ polytype in this study. 
The least energetically favorable polytype is $2H$ whose total energy is higher than $6H$ by 7.1~meV per SiC. 
We have found that, compared with other materials, SiC exhibits smaller energy difference among polytypes. This is derived from the balance of ionicity and covalency. The materials with dominantly ionicity prefer hexagonal structure. SiC is a exquisite material possessing a delicate balance of ionicity and covalency to exhibit hundreds of polytypes \cite{Kobayashi2}. 
As for the other materials, most stable structure of each material is $2H$-AlN, $2H$-BN, $2H$-GaN, $3C$-Si, and $3C$-C, respectively. 
The most stable structures in other materials are commonly observed in experiments. 

\begin{figure}
\includegraphics[width=0.9\linewidth]{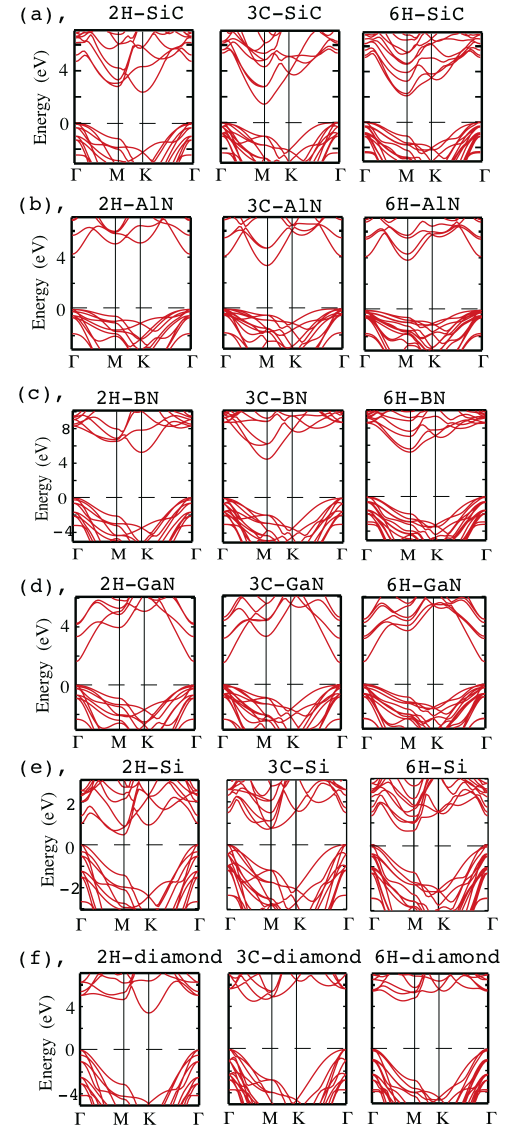}
\caption{(Color online). Band structures calculated by the GGA. The energy of the valence-band top is set to be 0. In these calculations, we adopted supercell calculations, so that the number of electrons is equal to each other for easy comparison and they have the same Brillouin zone. Note that $3C$ structures are also calculated in the hexagonal supercell, thus the $X$ point in cubic cell being folded to $M$ point.}
\label{bands}
\end{figure}

Next we have calculated electronic band structure for each material. The calculated results are shown in Fig.~\ref{bands}. Remark that we have adopted a unit cell of the $6H$ structure even for $2H$ and $3C$ structures to facilitate the comparison among the polytypes. 
From the figures, the valence bands of the three polytypes resemble each other in each material. The tiny differences are attributed to the difference of the symmetries by which degenerate states in the high-symmetry structure split. The valence-band top is located at $\Gamma$ point in all the polytypes in all the materials. In contrast, the conduction bands are qualitatively different among polytypes in spite of their structural similarity in the local atomic arrangement. In the SiC polytypes, the CBM is located at $K$ point in the $2H$-structure, whereas it is at $M$ point in the $3C$-, and $6H$-structure. The $X$ point in the cubic Brillouin zone (BZ) is folded on the $M$ point in the hexagonal BZ. Furthermore, the lowest conduction band in the $3C$ structure is isolated and shifts downwards substantially, making the band gap narrower by 0.7 - 0.9 eV than those in the $6H$ and $2H$ polytypes. The calculated energy bands for other compounds clearly show the same feature as in SiC, i.e., the CBM in the $3C$-AlN, $3C$-BN is located at $M$ point, whereas that in the $2H$-BN, and $2H$-diamond is located at the $K$ point.


The calculated and experimental band gaps for the polytypes are given in Table~\ref{bandgap}. Overall features of the calculated band-gap variation are in accord with the experimental values. It is clearly seen that the GGA underestimates energy gaps by about 50\% because of the shortcoming inherent in the GGA. If necessary, the quantitative description of the energy gaps is possible using more sophisticated schemes of the GW \cite{GW1, GW2, GWissue} for quasiparticle-self energy or HSE functional \cite{HSE, HSE04, HSE05, HSE06, matsushita4} for the exchange-correlation energy. Yet, the relative difference in the energy gap calculated by the GGA among the polytypes is well reproduced, i.e., calculated results show the band gap of the $3C$-SiC is smaller than that of the $2H$-SiC by 0.936 eV, which corresponds to the experimental one, 0.93 eV.

\begin{table} 
\begin{center}
\caption{
Calculated, $\epsilon_{\rm gap}$, and experimental, $\epsilon_{\rm expt.}$, energy gaps of the $2H$, $3C$ and $6H$ structures for various ${\it sp}^3$-bonded semiconductors. Experimental data are taken for SiC from Ref.~\cite{Harris}, for $2H$-AlN from Ref.~\cite{2H-AlN-bandgap}, for $3C$-AlN from Ref.~\cite{3C-AlN-bandgap}, for $3C$-BN from Ref.~\cite{3C-BN-bandgap}, for GaN, Si, and diamond from Ref.~\cite{3C-GaN}.
\label{bandgap}}
\scalebox{1.2}[1.2]{
\begin{tabular}{ccccc}
\hline
Materials & $\epsilon_{\rm gap}$ &$\epsilon_{\rm expt.}$  
\\
                  & (eV) & (eV)
\\ [2pt]  \hline  \\ [-4pt]
$2H$-SiC &2.355   (indirect)~ & 3.33 (indirect)
\\
$3C$-SiC &1.419   (indirect)~ & 2.40 (indirect)
\\
$6H$-SiC & 2.077   (indirect)~ & 3.10 (indirect)
\\ [2pt]  \hline  \\ [-4pt]
$2H$-AlN & 4.233   (direct)~ & 6.23 (direct) 
\\
$3C$-AlN &3.328   (indirect)~ & 5.34 (indirect) 
\\
$6H$-AlN &3.817   (indirect)~ & $-$ 
\\ [2pt]  \hline  \\ [-4pt]
$2H$-BN & 5.251   (indirect)~ & $-$
\\
$3C$-BN & 4.487   (indirect)~& 6.4   (indirect)
\\
$6H$-BN & 5.190   (indirect)~& $-$
\\ [2pt]  \hline  \\ [-4pt]
$2H$-GaN &1.622   (direct)~& 3.28   (direct)
\\
$3C$-GaN &1.489   (direct)~& 3.47   (direct)
\\
$6H$-GaN &1.533   (indirect)~&$-$
\\ [2pt]  \hline  \\ [-4pt]
$2H$-Si &0.477   (indirect)~&$-$
\\
$3C$-Si &0.660   (indirect)~& 1.17   (indirect) 
\\
$6H$-Si &0.639   (indirect)~&$-$
\\ [2pt]  \hline  \\ [-4pt]
$2H$-diamond ~~&3.406   (indirect)~&$-$
\\
$3C$-diamond ~~&4.246   (indirect)~& 5.48   (indirect)
\\
$6H$-diamond ~~&4.521   (indirect)~&$-$\\
\hline
\end{tabular}
}
\end{center}
\end{table}

From the Table~\ref{bandgap}, it has been found that the large band-gap variation is not limited to the SiC polytypes. For AlN and BN, the energy gap decreases substantially in the $3C$ structures by 0.9 eV and 0.8 eV, respectively. In the case of AlN, the CBM at $M$ point shifts downwards substantially, so that the transition between the direct gap in the most stable $2H$-structure and the indirect gap in the metastable $3C$-structure takes place. This result gives good agreement with the observed experimental facts. 
In contrast, for the diamond polytypes, the band-gap decrease can be seen not at the $3C$-structure, but at the $2H$-structure: the energy gap varies from 4.521 eV in the $6H$-structure to 3.406 eV in the $2H$-structure.

\subsection{Floating states in $3C$ structure} \label{result_3C}

\begin{figure}
\includegraphics[width=0.5\textwidth]{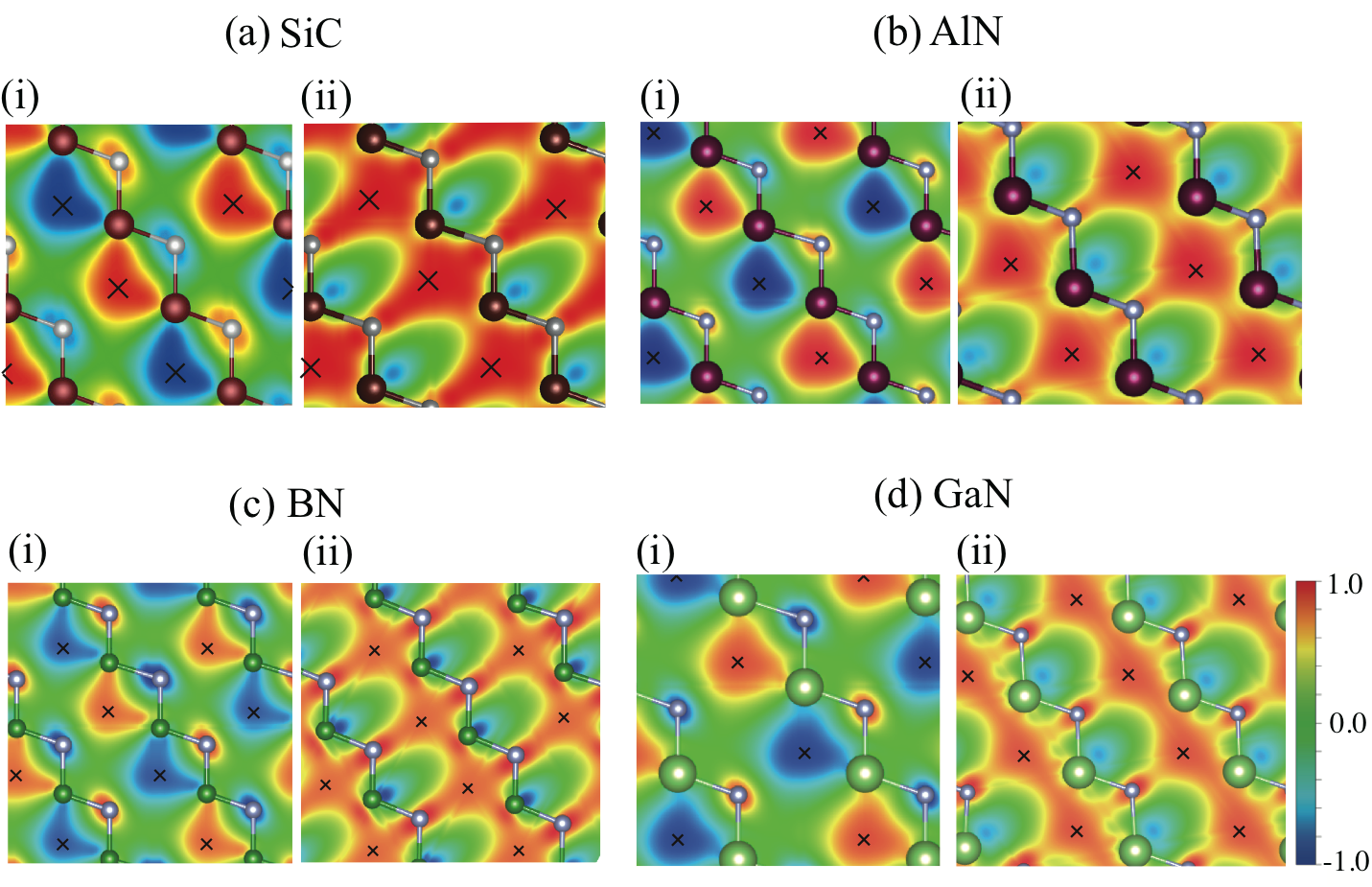}  
\caption{(Color online) Contour plots of the calculated Kohn-Sham (KS) orbitals of the conduction-band minimum at $M$ point for $3C$-SiC (a), $3C$-AlN (b), $3C$-BN (c), and $3C$-GaN (d) on the $\left(0{\bar 1}1\right)$ (left panel labeled (i)) and the $\left(110\right)$ (right panel labeled (ii)) plane. The $M$ point which we discuss corresponds to $X = (0,0,2\pi/a_0)$ in cubic BZ. The mark 'X' depicts the tetrahedral ($T_d$) interstitial sites surrounded by cations. In Fig. (a), brown(large) and white(small) balls depict silicon and carbon atoms, respectively. In Fig. (b), brown(large) and white(small) balls depict aluminum and nitrogen atoms, respectively. In fig. (c), green and white balls are boron and nitrogen atoms, respectively, in Fig. (d), green and white is gallium and nitrogen atoms, respectively.}
\label{3C-floating}
\end{figure}

\begin{figure}
\begin{center}
\includegraphics[width=0.4\textwidth]{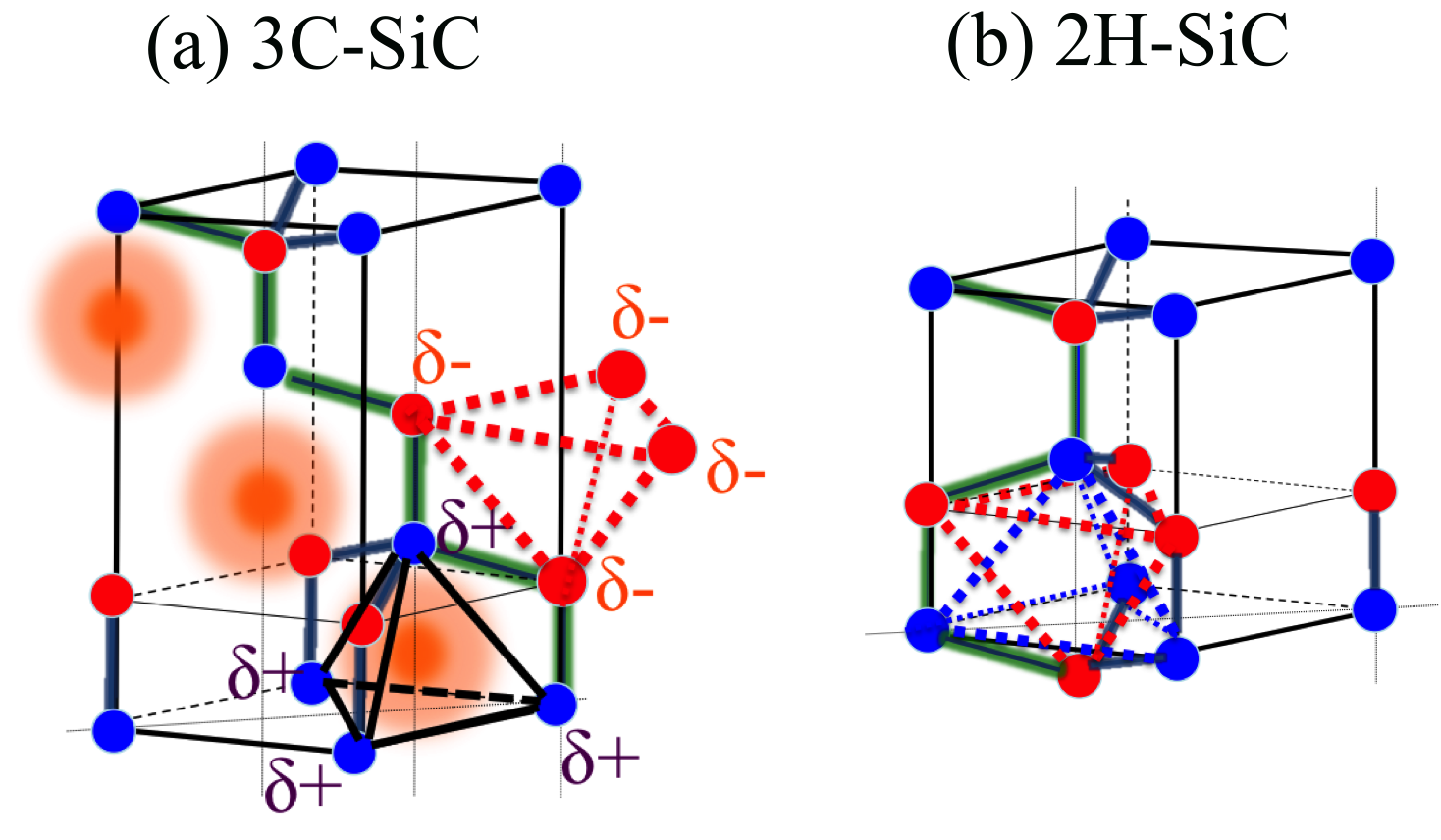}  
\end{center}
\caption{(Color online) Sketches of two tetrahedral ($T_d$) interstitial sites in the $3C$-polytypes (a), and the $2H$-polytypes (b): One is surrounded by 4 cations and the other is by 4 anions. The blue balls represent cations, and the red ones anions. In the $3C$-polytype (a), the cation-surrounded interstitial site is spatially separated from the anion one. On the other hand, they overlap each other in the $2H$-polytype.}
\label{SiC-interstitial}
\end{figure}

We discuss the microscopic mechanism of the band-gap variation in this subsection. As we have clarified in the previous papers \cite{matsushita1,matsushita2,matsushita3}, continuum-state like character at CBM in SiC polytypes plays important roles in the band-gap variation. As shown in Fig.~\ref{3C-floating} (a), 
the CBM of $3C$-SiC extends (or $floats$) in internal channel cavity without atomic orbital character, i.e., $\langle 110 \rangle$ channels.
Si atoms are positively charged because of the differences in electronegativity between Si and C atom in SiC crystals (see Fig.~\ref{SiC-interstitial}(a)). Thus, this charge transfer causes the electro-static potential at the tetrahedral ($T_{d}$) interstitial sites surrounded by 4 Si atoms lower. These $T_d$ interstitial sites construct the $\langle 110 \rangle$ channels where the floating state extends having the maximum amplitude at $T_d$ sites. This lowering of electro-static potential at $T_{d}$ interstitial sites shifts the energy level of the floating state downwards. 

First we discuss the character of the CBM in other $sp^3$-bonded semiconductors in 3$C$ structure in Fig.~\ref{3C-floating}. 
As clearly seen in Fig.~\ref{3C-floating}, the Kohn-Sham (KS) orbitals at the CBM at $M$ point in $3C$ structure on $\left(0\bar{1}1\right)$ 
plane obtained in the GGA calculations are similar to that of SiC, indicating that the CBMs at $M$ point in other $sp^3$-bonded semiconductors are also floating states extending in $\langle 110 \rangle$ channels. These structures have similar channel features as SiC: $T_d$ interstitial sites surrounded by cations form $\langle 110 \rangle$ channels, rendering the energy level of floating states lower. 


In contrast, elemental semiconductors, such as Si and diamond, exhibit no such band-gap variation in $3C$ structure. It is because there is no charge transfer unlike the compound ones. This fact makes no potential lowering at $T_d$ interstitial sites, causing no band-gap narrowing in elemental semiconductors. 

Next we discuss why band-gap variations are not seen in $2H$ structure. As mentioned above, charge transfer plays important roles in the substantial band-gap decrease in the $3C$ structures. On the other hand, in the $2H$-structure such a cation-surrounded channel is absent. The internal space surrounded by cations overlaps considerably with that by anions in the $2H$ structure [See Fig.~\ref{SiC-interstitial}]. The cation-surrounded interstitial site is very close to the anion-surrounded one with the separation of $d/3$, where $d$ is the bond length between silicon and carbon atoms. In fact, the electro-static potential at the cation-surrounded interstitial sites is almost the same as that at anion-surrounded ones within 0.1 eV in the case of SiC. Therefore, in the $2H$ structures, charge transfer doesn't cause the static potential lowering, leading to no band-gap variation. 

\subsection{Floating states in $6H$ structure} \label{result_6H}
\begin{figure}
\begin{center}
\includegraphics[width=0.4\textwidth]{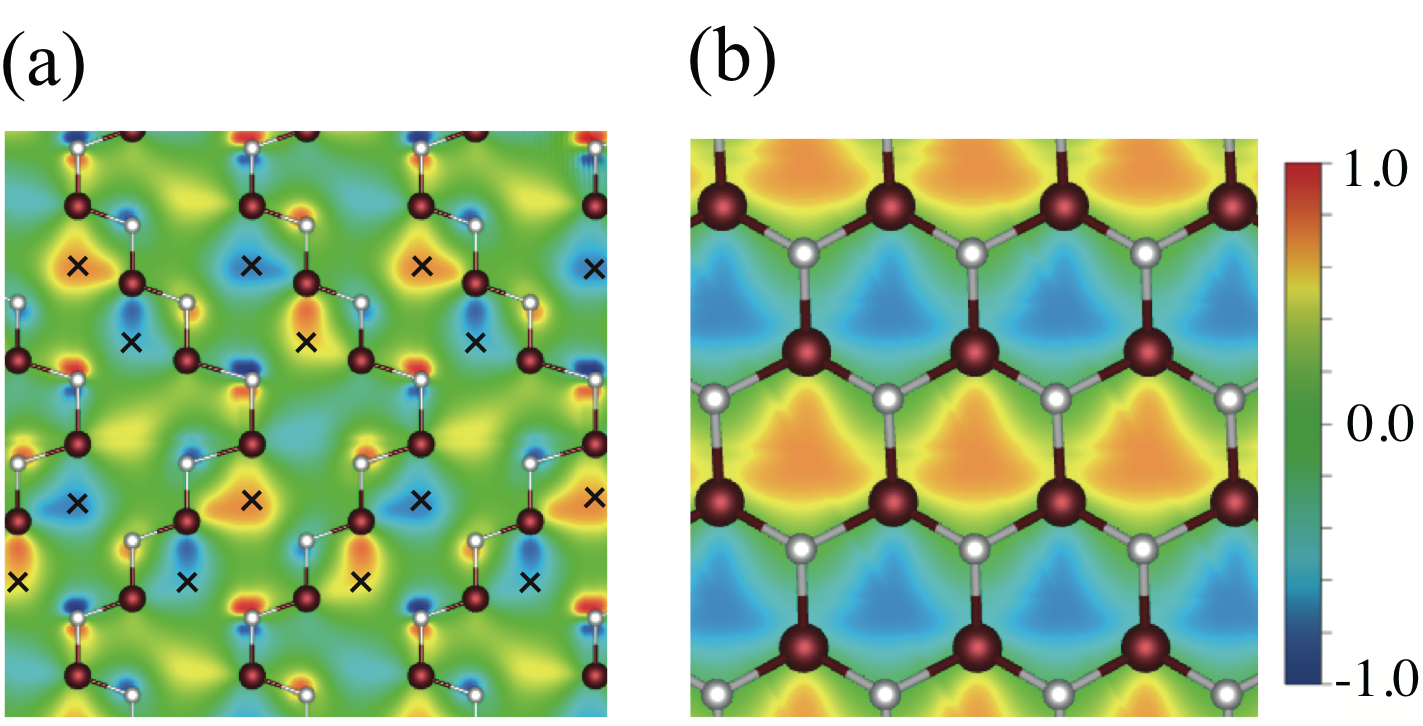}  
\end{center}
\caption{(Color online) Contour plots of the calculated Kohn-Sham (KS) orbitals of the conduction-band minimum at $M$ point for $6H$-SiC on $\left(11\bar{2}0\right)$ plane in (a), and $\left(0001\right)$ in plane (b). The brown and white balls depict Si atoms and C atoms, respectively. The mark 'X' represents the tetrahedral ($T_d$) interstitial sites surrounded by Si atoms.}
\label{6H-SiC}
\end{figure}

In this subsection we discuss the electronic structure at CBM in $6H$ structure. As mentioned above, in the $3C$ structures, the $\left<110\right>$ channels with infinite length play important roles in the variations in energy gaps. Similar channel structure is seen also in $6H$ structure. There is channels with the length of about $7 a_0 / 2 \sqrt{2}$ along $\langle \bar{2}201 \rangle$ which is slanted relative to $\langle 0001\rangle$ direction with $a_0$ a lattice constant. 
The calculated KS orbital at the CBM of SiC is shown in Fig.~\ref{6H-SiC}, where the wavefunction has the maximum amplitude at the tetrahedral $T_d$ interstitial sites, and $floats$ in the finite-length channels. Due to quantum confinement of the wavefunction, however, the kinetic energy at the CBM is greater than that in $3C$ structure and the band gap of $6H$ structure becomes wider [See Fig.~\ref{bands} and Table.~II] by 0.66 eV. The relations between the channel length and band gap is clearly shown in Fig.~\ref{seibun}. 
Similar tendency is observed also in other $sp^3$ compound semiconductors. AlN, and BN in $6H$ structure exhibit 0.49 eV, and 0.7 eV wider band gap than that in $3C$, respectively.

\subsection{Floating states in $2H$ structure}\label{result_2H}
\begin{figure}
\begin{center}
\includegraphics[width=0.45\textwidth]{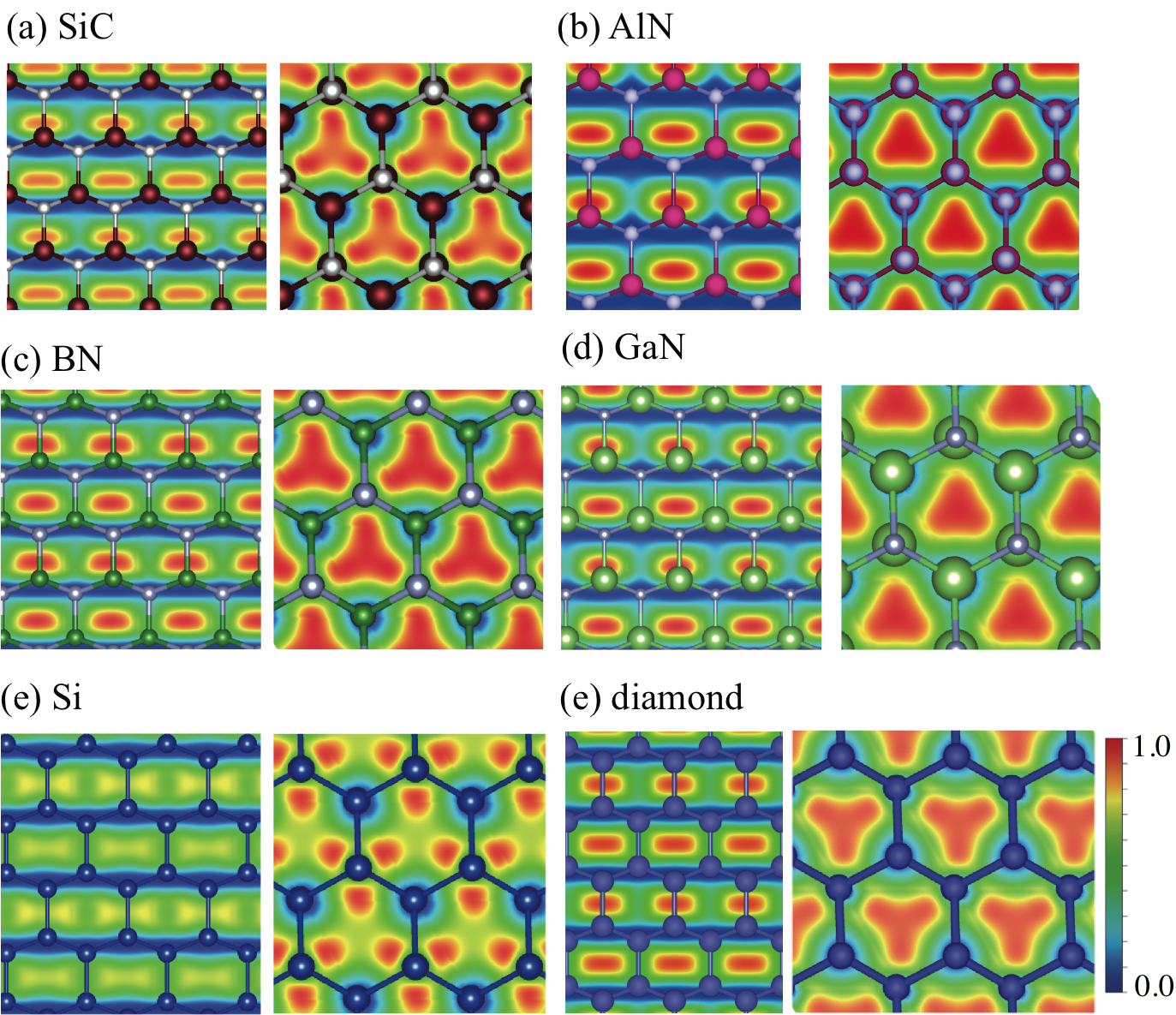}  
\end{center}
\caption{(Color online) Contour plots of the calculated Kohn-Sham(KS) orbitals of the conduction-band minimum at $K$ point for $2H$-SiC (a), $2H$-AlN (b), $2H$-BN (c), $2H$-GaN (d), and $2H$-Si on $\left(1\bar{1}00\right)$ (left panel) and $\left(0001\right)$ (right panel) planes. In Fig.~(a), the brown and white balls depict silicon and carbon atoms, in Fig.~(b), the pink and sky blue balls are aluminum and nitrogen atoms, in Fig.~(c), green and white balls are boron and nitrogen atoms, in Fig.~(d), green and white are gallium and nitrogen atoms, respectively, and in Fig.~(e), the blue balls are silicon atoms. }
\label{2H-K}
\end{figure}

In this subsection we discuss the electronic structure at CBM in $2H$ structure, and give an explanation why diamond exhibits smallest band gap at $2H$ structure. 
The floating state at $M$ point in the $3C$-polytypes is distributed along the $\langle 110 \rangle$ channel which is slanted relative to $\langle 111\rangle$ direction. In the $2H$ structure, such a slanted channel is absent. Instead, there are channels along $\langle 11\bar{2}0 \rangle$ and $\langle 0001 \rangle$ directions. We have found that the CBM at $K$ point in the $2H$ structure floats in the $\langle 11\bar{2}0 \rangle$  channel  in SiC case (Fig.~\ref{2H-K}(a)). This floating state is distributed solely in the $\langle 11\bar{2}0 \rangle$ channel with its phase changing consecutively by $\exp (i 2 \pi / 3)$, thus avoiding the atomic sites on (0001) planes. It is also found that the floating state is distributed closer to the planes of positively-charged Si atoms to gain the electro-static energy. The phase change of the floating orbital along this channel is compatible with the symmetry of the Bloch state at $K$ point. 

\begin{figure}
\begin{center}
\includegraphics[width=0.45\textwidth]{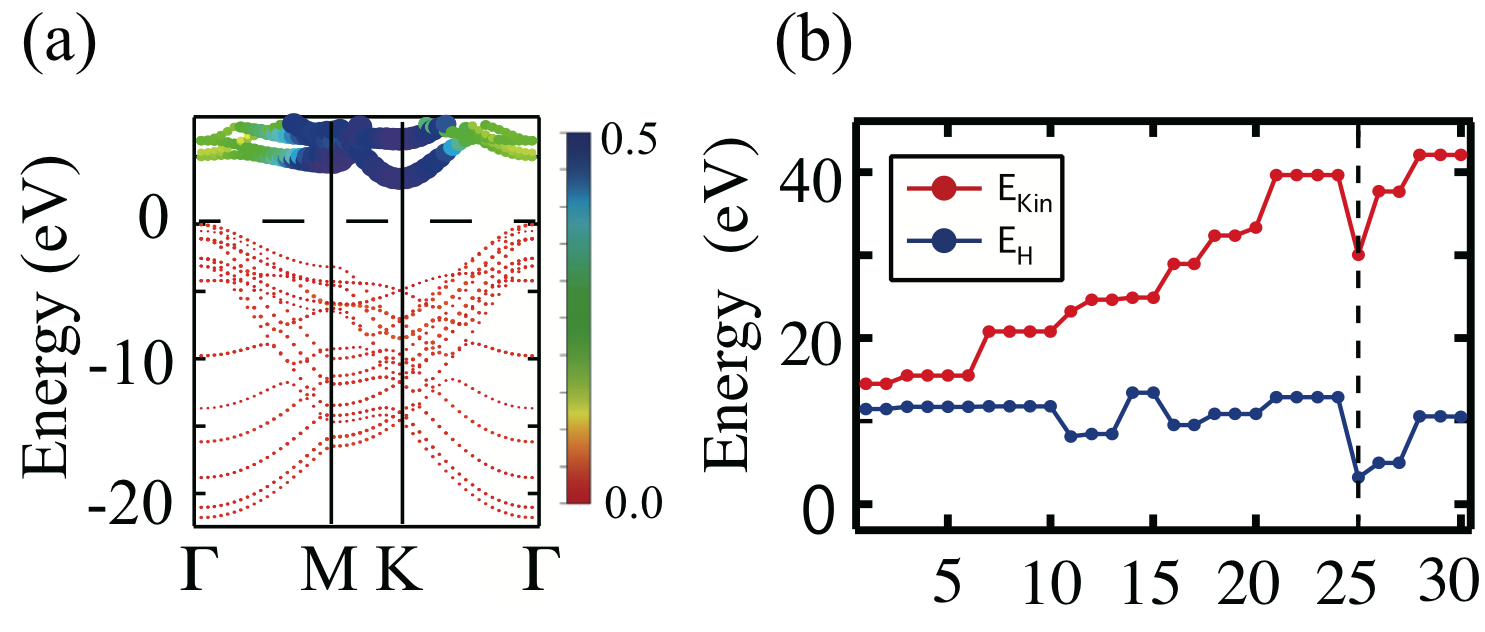}  
\end{center}
\caption{(Color online) (a) Residual norms of the wavefunctions of the energy bands of 2$H$-diamond. The residual norms are represented by the color and the size of the dots. The energy of the valence-band top is set to be 0. The residual norm which is a measure of the floating nature is calculated in the following procedure: From the pseudo-atomic orbitals $\{\phi_i^{\rm isolated}\}$ of isolated carbon atoms, we have composed orthonormal basis set $\{\phi_i^{\rm atom}\}$ with the Gram-Schmidt orthonormalization. Then we have calculated the squared residual norm, which is defines as $\left| \left|\phi_{n{\bf k}}\right> -\sum_i \left|\phi_i^{\rm atom}\right>\left<\phi_i^{\rm atom}|\phi_{n{\bf k}}\right> \right|^2$ for each band n at ${\bf k}$ point. (b) Energy analyses of KS orbitals in $2H$-diamond. The kinetic-energy contribution $\epsilon_{\it kin}$ and the Hartree-energy contribution $\epsilon_{\rm H}$ to the orbital energy of each KS state for $K$ point. The abscissa represents the $i$th KS state from the valence-band bottom and the 25th state is the conduction-band minimum.}
\label{2H-diamond-CBM}
\end{figure}


We then expect that the existence of the floating states is common to $2H$ structure in most $sp^3$-bonded semiconductors. We have therefore examined $2H$-AlN, $2H$-BN, $2H$-GaN, $2H$-Si, and $2H$-diamond. Fig.~\ref{2H-K} shows the CBM at $K$ point of them. We clearly see the floating states in all materials. 
Most remarkable case is diamond. Diamond polytypes show substantial band-gap decrease in the $2H$ structure. Fig.~\ref{2H-K}(e) shows the KS orbital at the CBM of the $2H$-diamond. The KS orbital is also distributed not near atomic sites, but $floats$ in internal space. 
Fig.~\ref{2H-diamond-CBM}(a) shows the residual norm of wavefunction after the projection to the $s$-, and $p$-atomic orbitals. As much as 0.47 are floating character at the CBM in the $2H$-diamond, while it is only 0.04 at the valence-band top. This fact shows that this KS orbital at the CBM does not consist of atomic orbitals neither. The maximum amplitude is on the axis of the $\left<0001\right>$ channels, and the floating state is distributed in the horizontal channels. We have found that the floating state induces band-gap variations also in this system. 

In order to clarify the reason for the energy gain of the floating state, we show the energy analyses of KS orbitals at $K$ point in Fig.~\ref{2H-diamond-CBM}(b). According to the figure, floating state reveals the kinetic energy gain by extending in the channels broadly. Another energy gain is Coulomb energy gain, because the floating state is distributed far from atomic nuclei, which core electrons are distributed near around. In addition, we have found that the electro-static potential from the ions energy gain plays important roles in the decrease of the energy gap in the $2H$-diamond. 
In fact, the local atomic structure around the interstitial sites is different between the $3C$ and $2H$ structure. The interstitial site in $2H$ structure is surrounded by six nearest neighbor atoms, and six next nearest neighbor atoms. In contrast, the interstitial site in the $3C$ polytype is surrounded by four nearest neighbor, and three next neighbor atoms. That is, in the $2H$ polytypes, the number of neighbor atoms around the interstitial sites is larger compared with other polytypes. 
This structural difference makes the electro-static potential at the interstitial sites in the $2H$-diamond lower than that in the $3C$-diamond by 0.589 eV from the DFT calculation. That value corresponds to the band-gap variation in the $2H$ structure, 0.8 eV smaller than $3C$. 

\section{Conclusions}
We have performed electronic structure calculations to explore the band gap dependence on polytypes for $sp^3$-bonded semiconducting materials, i.e., SiC, AlN, BN, GaN, Si, and diamond. We have found that band-gap variation is common in $sp^3$-bonded semiconductors; SiC, AlN, and BN exhibit smallest band gaps in $3C$ structure, whereas diamond does in $2H$ structure. We have also clarified that the microscopic mechanism of the band-gap variations is attributed to peculiar electron states $floating$ in internal channel space at the conduction-band minimum (CBM), and that channel length and electro-static potential in channel space affect the energy level of the floating states; In compound semiconductors, charge transfer causes the elecro-static potential in channel lower in $3C$ structure, while elemental semiconductors show lower electro-static potential in $2H$ rather than in $3C$.

\section{Acknowledgements}
\begin{acknowledgments}
This work was supported by the Grant-in-Aid for Young Scientists (B) conducted by MEXT, Japan, under Contract No. 93002181. This research (in part) used computational resources of COMA provided by Interdisciplinary Computational Science Program in Center for Computational Sciences, University of Tsukuba, and the Supercomputer Center at the Institute for Solid State Physics, The University of Tokyo.
\end{acknowledgments}

\section{Appendix}
Here we show the specific values of band gaps of SiC polytypes in Table.~\ref{Appendix}.

\begin{table} 
\begin{center}
\caption{
Calculated energy gaps, $\epsilon_{\rm gap}$, of various SiC polytypes. 
\label{Appendix}}
\scalebox{1.2}[1.2]{
\begin{tabular}{ccccc}
\hline
Polytypes & $\epsilon_{\rm gap}$ &Channel length  
\\
                  & (eV) & 
\\ [2pt]  \hline  \\ [-4pt]
3$C$ structure\\
ABC &2.25& $\infty$
\\ [2pt]  \hline  \\ [-4pt]
4$H$ structure\\
ABCB&3.18&3
\\ [2pt]  \hline  \\ [-4pt]
5$H$ structure\\
ABCBC&2.85&5
\\ [2pt]  \hline  \\ [-4pt]
6$H$ structures\\
ABCBCB&3.35&3 
\\
ABCACB&2.92&4 
\\ [2pt]  \hline  \\ [-4pt]
8$H$ structures\\
ABCABACB&2.68&5
\\
ABCABCAB&2.51&8
\\
ABCACBAB&3.00&4
\\
ABCACBCB&3.00&4
\\
ABCBABAB&3.30&3
\\
ABCBCBAB&3.18&3
\\ [2pt]  \hline  \\ [-4pt]
10$H$ structures\\
ABCABCABAB&2.40&8
\\
ABCABCABCB&2.39&9
\\
ABCABCBACB&2.57&6 
\\
ABCACBCACB&2.95&4
\\
ABCBCACBCB&3.33&3
\\
ABCBCBCBCB&3.31&3
\\ [2pt]  \hline  \\ [-4pt]
12$H$ structures\\
ABACABCBCACB&2.96&4\\
ABCABACABACB&2.68&5\\
ABCABACBCACB&2.76&5\\
ABCABCABCACB&2.36&10\\
ABCABCACBACB&2.49&7\\
ABCACACACACB&2.81&4\\
ABCBCBCBABAB&3.28&3\\
\hline
\end{tabular}
}
\end{center}
\end{table}

\end{document}